\documentstyle[aps]{revtex}
\renewcommand{\theequation}{\arabic{section}.\arabic{equation}}

\newcommand{\be}{\begin{equation}}
\newcommand{\ee}{\end{equation}}
\newcommand{\eqn}{{eq.~}}
\newcommand{\eqns}{{eqs.~}}

\newcommand{\Eqns}{{Eqs.~}}
\def\ba{\begin{eqnarray}}
\def\ea{\end{eqnarray}}

\def\o{\omega}
\def\f{\frac}
\def\({\left(}
\def\){\right)}
\def\ls{\left[}
\def\rs{\right]}

\def\et{{\epsilon}_{\tau}}
\begin{document}

\title{\vspace*{1.5in}Vacuum polarization in the Schwarzschild spacetime
and dimensional reduction}
\author{R. Balbinot$^{(a)}$\thanks{Electronic address: balbinot@bo.infn.it},
A. Fabbri$^{(a)}$\thanks{Electronic address: fabbria@bo.infn.it},
P. Nicolini$^{(a)}$\thanks{Electronic address: nicolini@bo.infn.it},
V. Frolov$^{(b)}$\thanks{Electronic address: frolov@phys.ualberta.ca},
P. Sutton$^{(c)}$\thanks{Electronic address:
psutton@gravity.phys.psu.edu}, and
A. Zelnikov$^{(b,d)}$\thanks{Electronic address: zelnikov@phys.ualberta.ca}
}
\address{$^{(a)}$Dipartimento di Fisica dell'Universit\`a di Bologna and INFN
sezione di Bologna, Via Irnerio 46,  40126 Bologna, Italy\\
$^{(b)}$Theoretical Physics Institute, Department of Physics,
University of Alberta,
Edmonton, AB, Canada T6G 2J1\\
$^{(c)}$Department of Physics, The Pennsylvania State University,  
State College, PA, USA 16802-6300\\
$^{(d)}$P.N.~Lebedev Physics Institute,
Leninsky pr. 53, Moscow  117924, Russia}
\maketitle

\bigskip

\begin{abstract}
A massless scalar field minimally coupled to gravity and
propagating in the Schwarzschild spacetime is considered. After
dimensional reduction under spherical symmetry the resulting 2D
field theory is canonically quantized and the renormalized
expectation values $\left< T_{a b}\right>$ of the relevant
energy-momentum tensor operator are investigated. Asymptotic
behaviours and analytical approximations are given for $\left<
T_{a b}\right>$ in the Boulware, Unruh and Hartle-Hawking states.
Special attention is devoted to the black-hole horizon region
where the WKB approximation breaks down.

\medskip \noindent
PACS number(s): 04.62.+v, 11.10.Gh, 11.10.Kk
\end{abstract}
\bigskip

%%%%%%%%%%%%%%%%%%%%%%%%%%%%%%%%%%%%%%%%%
\section{Introduction}

In Quantum Field Theory the dimensional reduction of a system obeying
some symmetries, such as spherical symmetry, is obtained by decomposing 
the field operators in harmonics in the symmetrical subspace. 
In the case of spherical symmetry, decomposing in terms of spherical 
harmonics effectively reduces a $4D$ theory to a set of $2D$ theories
characterized by different values of the angular momentum. %\\ \noindent

Two-dimensional theories are often regarded as useful tools for inferring 
general features of systems whose behaviour is sophisticated and
difficult to analyze in the physical $4D$ spacetime. In some spherically
symmetric systems the main physical effects come from the  
``s-wave sector'' -- the $l=0$ mode. 
Truncation of higher momentum modes is then obtained 
by integrating over the ``irrelevant'' angular variables.
This is the spirit which pervades most of the vast literature on 2D black
holes, though this s-wave approximation is not always accurate enough.
These  models are believed to describe the 
s-wave sector of physical $4D$ black holes.  %\\ \noindent

Within this perspective, a model of 
$2D$ conformally invariant matter fields interacting with 
$2D$ dilaton gravity has attracted considerable interest
recently.  The action for this theory is 
\be
S=-\frac{1}{2} \int d^2 x\sqrt{-g}e^{-2\phi}g^{a
b}\partial_a\varphi\partial_b\varphi \, ,
\label{act}
\ee
where $\varphi$ is the scalar field, $\phi$ the dilaton, $g_{a b}$ the
2D background metric and $a,b=1,2$. %\\ \noindent

The reason for this interest lies in the following: 
the action (\ref{act}) can be obtained by the dimensional 
reduction of the 4D action for a massless
scalar field minimally coupled to 4D gravity, 
\be
S^{(4)}=-\frac{1}{8\pi}\int d^4
x\sqrt{-g^{(4)}}g^{\mu\nu}\partial_{\mu}\varphi\partial_{\nu}\varphi \, ,
\label{foact}
\ee 
under the assumption of spherical symmetry.
Decomposing the 4D spacetime as %follows 
\be 
ds^2 =
g_{\mu\nu}^{(4)}dx^{\mu}dx^{\nu}= g_{ab} dx^a dx^b +
e^{-2\phi(x^a)}d\Omega^2 \>,  \label{constr} 
\ee 
where $d\Omega^2$ is the metric on the unit two-sphere, 
one obtains the 2D action
(\ref{act}) by inserting the decomposition (\ref{constr}) into the
action (\ref{foact}), imposing $\varphi=\varphi(x^a)$, and
integrating over the angular variables. Therefore the model based
on the action (\ref{act}) seems more appropriate for discussing 
the quantum properties of black holes 
in the s-wave approximation  
than other 2D models based on the Polyakov action
(describing a minimally coupled 2D massless scalar field), 
whose link with the real 4D world is missing. 
For this reason the efforts of many authors were
devoted to finding the effective action which describes at the
quantum level the above 2D dilaton gravity theory ($\!$\cite{MuWiZe}; see
also \cite{KuVa} and \cite{NoOd}) . This effective
action, once derived, would allow one to go beyond the fixed
background approximation usually assumed in the studies of the
quantum black-hole radiation discovered by Hawking \cite{Hawk}. Such an
effective action will give in fact $\left< T_{ a b}\right>$ for an
arbitrary 2D spacetime which could then be used to study
self-consistently, within this 2D approach, the backreaction of an
evaporating black hole, its evolution, and its final fate.
Unfortunately the effective actions so far proposed for the model
of \eqn (\ref{act}) have serious problems in correctly reproducing 
Hawking radiation even in a fixed Schwarzschild spacetime (see
the discussion in Ref. \cite{BaFa}; see also \cite{KuVa} for a
different point of view). In any case before embarking on
ambitious backreaction calculations and taking seriously puzzling
results (such as antievaporation \cite{BuRa}) one should 
check for any candidate
of the effective action that leads, at least for the Schwarzschild
black hole, to the correct results. But what are the exact $\left<
T_{a b}\right>$ for a scalar field described by the action
(\ref{act}) propagating in a 2D Schwarzschild spacetime that the
relevant effective action should predict? The aim of this paper is
to partially answer this question. %\\ \noindent 

By standard canonical
quantization we will be able to give the asymptotic (at infinity
and near the black hole horizon) values of $\left< T_{a b}\right>$
in the three quantum states relevant for a field in the
Schwarzschild spacetime, namely: the Boulware state (vacuum
polarization around a static star), the Unruh state (black hole
evaporation), and the Hartle-Hawking state (black hole in thermal
equilibrium). We will also obtain approximate analytical
expressions for $\left< T_{a b}\right>$ for every value of the
radial coordinate. Any effective action for the model of \eqn 
(\ref{act}) which is unable to predict at least the above
asymptotic values of $\left< T_{a b}\right>$ is incorrect (or
better incomplete) and any result based on it has no physical
support.

\setcounter{equation}{0}
\section{$\left< T_{a b}\right>$: asymptotic behaviour}
Our main goal is the evaluation of the renormalized expectation values
of the stress tensor operator for the scalar field $\varphi$ whose
dynamics are given by the action (\ref{act}). Here we will be interested in
the asymptotic values (at infinity and near the horizon). The following
derivation is just a readaptation to our model of section VI of the
seminal paper by Christensen and Fulling \cite{ChFu} to which we refer the
reader (see also \cite{Cand}). %\\ \noindent
 
The classical stress tensor is defined as
\be
T_{ab} = - \frac{2}{\sqrt{-g}}\frac{\delta S}{\delta
g^{ab}} \> ,
\label{emt}
\ee
hence from \eqn (\ref{act})
\be T_{ab}=
e^{-2\phi} \left[\partial_a{\varphi}\partial_b{\varphi} -
\frac{1}{2}g_{ab}\left(
\nabla\varphi\right)^2 \right] \>.
\ee
The scalar field obeys the field equation
\be
\nabla^a\( e^{-2\phi}{\nabla}_a\varphi\)=0 \> .
\label{emot}\ee
The quantum field operator $\hat\varphi$ is then expanded on 
a basis $\{ u_j \}$ for the solution of \eqn (\ref{emot}) in 
terms of annihilation and creation operators, 
\be
\hat \varphi =\sum_j\left( \hat{a}_j u_j + \hat{a}_j ^{\dagger}
u_j ^{\ast} \right) \, ,
\ee
and computing the mean value $\left<0|T_{ab}|0\right>$ we have
\be
\left< T_{ab}\right>=\sum_j T_{ab}\left[
u_j,u_j^{\ast}\right]\>, 
\ee
where
\be
T_{ab} \left[ u_j , u_j ^{\ast}
\right] =  e^{-2\phi} \left\{ Re\left[\left(\nabla_a u_j
\right)\left(\nabla_b u_j ^{\ast}\right)\right]- (1/2)g_{ab}{|\nabla
u_j|}^2\right\}.
\ee
Taking as the background geometry the exterior Schwarzschild solution
\be
ds^2=-(1-2M/r)dt^2 + (1-2M/r)^{-1}dr^2 ,\ \ \ \phi=-\ln r \label{sclo} \> , 
\ee
one finds that a set of normalized basis functions of the field equation
(\ref{emot}) is given by
\be
{\stackrel{\rightarrow}{u}}_{w}(x)=\frac{1}{\sqrt{4\pi
w}}\frac{{\stackrel{\rightarrow}{R}}(r;w)}{r}e^{-iwt},
\label{u1}
\ee
\be
{\stackrel{\leftarrow}{u}}_{w}(x)=\frac{1}{\sqrt{4\pi
w}}\frac{{\stackrel{\leftarrow}{R}}(r;w)}{r}e^{-iwt}  ,
\label{u2}
\ee
where the radial functions $R(r;w)$ satisfy the differential
equation
\be
-\frac{d^2R}{d{r^{\ast}}^2}+\left( 1-2M/r \right)\left[
\frac{2M}{r^3}\right]R-w^2R=0 \, ,
\label{barr}
\ee
and $r^{\ast}$ is the Regge-Wheeler coordinate
\be
r^{\ast}=r+2M\ln(r/2M -1)\>.
\ee
Exact solutions of \eqn (\ref{barr}) are not known; however, one can 
find their asymptotic behaviour near the horizon,
\ba 
 & & {\stackrel{\rightarrow}{R}}\sim e^{iwr^{\ast}}+
     {\stackrel{\rightarrow}{A}}(w)\, e^{-iwr^{\ast}} , \nonumber \\
 & & {\stackrel{\leftarrow}{R}}\sim \stackrel{\leftarrow}{B}
     (w)e^{-iwr^{\ast}} \label{ho} , 
\ea 
and at infinity, 
\ba 
 & & {\stackrel{\rightarrow}{R}}\sim \stackrel{\rightarrow}{B}(w)
     e^{iwr^{\ast}} , \nonumber \\ 
 & & {\stackrel{\leftarrow}{R}}\sim
     e^{-iwr^{\ast}} + {\stackrel{\leftarrow}{A}}(w)\, e^{iwr^{\ast}} . 
     \label{in}
\ea 
$A$ and $B$ are the reflection and transmission
coefficients (see Ref.  \cite{De}). %\\ \noindent

The $\left< T_{a b}\right>$ calculated for these modes corresponds 
to the so-called Boulware vacuum:
\be
\left< B| T_{a}^{\ b}
|B\right>_{unren}=\int_0 ^{\infty} dw\,\left\{ T_{a}^{\ b}  \left [
{\stackrel{\leftarrow} {u}}_w , {\stackrel{\leftarrow}{u}}_w
^{\ast} \right] +  T_{a}^{\ b}  \left[
{\stackrel{\rightarrow} {u}}_w , \stackrel{\rightarrow}{u}_w
^{\ast} \right]\right\}\> . 
\ee
For the Unruh vacuum we have \be \left< U| T_{a}^{\ b}
|U\right>_{unren}=\int_0 ^{\infty} dw\,\left\{ T_{a}^{\ b}  \left [
{\stackrel{\leftarrow} {u}}_w , {\stackrel{\leftarrow}{u}}_w
^{\ast} \right] + \coth{\left( 4\pi Mw\right)}\, T_{a}^{\ b}  \left[
{\stackrel{\rightarrow} {u}}_w , \stackrel{\rightarrow}{u}_w
^{\ast} \right]\right\}\>, \ee whereas for the Hartle-Hawking state
\be \left< H|T_{a}^{\ b}|H\right>_{unren}= \int_{0}^{\infty}dw \coth{(
4\pi
Mw)}\, \left\{
T_{a}^{\ b}\left[
\stackrel{\leftarrow}{u}_w, \stackrel{\leftarrow}{u}_w^{\ast}\right]\>
+T_{a}^{\ b}\left[
\stackrel{\rightarrow}{u}_w, \stackrel{\rightarrow}{u}_w^{\ast}\right]\>
\right\}.\ee 
As they stand
these expressions are ill-defined and need to be regularized.
However, taking into account the regularity of  
the renormalized expectation values $\left< H|T_{a
b}|H\right>$ on the horizon and the vanishing of $\left< B|T_{a
b}|B\right>$ as $r\to\infty$ , some asymptotic expressions can be
obtained without recursion to any regularization procedure.  For
example for $r\to\infty$ we can write \ba \lim_{r\to \infty}
\left<H|T_{a}^{\ b}|H\right> &=& \lim_{r\to\infty} 
(\left<H|T_{a}^{\ b}|H\right>- \left<B|T_{a}^{\ b}|B\right>)=
\lim_{r\to\infty}
(\left<H|T_{a}^{\ b}|H\right>- \left<B|T_{a}^{\ b}|B\right>)_{unren}
\nonumber \\ &=& \lim_{r\to\infty} 2\int_{0}^{\infty}
\frac{dw}{e^{8\pi Mw}-1}\left\{ T_{a}^{\ b} \left[
\stackrel{\rightarrow}{u}_w,{\stackrel{\rightarrow}{u}}_w^{\ast}\right]
+ T_{a}^{\ b} \left[
\stackrel{\leftarrow}{u}_w,{\stackrel{\leftarrow}{u}}_w^{\ast}\right]\
\right\}\>.\label{nb}\ea 
Similarly for the leading term at $r\to 2M$ we have \be
\lim_{r\to 2M}\left< B|T_{a}^{\ b}|B\right>\sim\lim_{r\to 2M} (\left<
B|T_{a}^{\ b}|B\right>- \left< H|T_{a}^{\ b}|H\right>)= \lim_{r\to
2M}(\left< B|T_{a}^{\ b}|B\right>- \left< H|T_{a}^{\ b}|H\right>)_{unren}
\>.\label{vv}
\ee 
For the Unruh vacuum we have 
\ba \lim_{r\to 2M}
\left<U|T_{a}^{\ b}|U\right>&\sim&\lim_{r\to 2M} 
(\left<U|T_{a}^{\ b}|U\right>- \left<H|T_{a}^{\ b}|H\right>)  =\lim_{r\to
2M}
(\left<U|T_{a}^{\ b}|U\right> - \left<H|T_{a}^{\ b}|H\right>)_{unren}\nonumber
\\ & & = \lim_{r\to 2M} \left\{
-2\int_{0}^{\infty}\frac{dw}{e^{8\pi Mw}-1} T_{a}^{\ b}\left[
\stackrel{\leftarrow}{u}_w,
\stackrel{\leftarrow}{u}_w^{\ast}\right]\right\} \> , \label{uv}
\ea
and 
\ba \lim_{r\to\infty}\left< U|T_{a}^{\ b}|U\right>&=&\lim_{r\to\infty}
(\left< U|T_{a}^{\ b}|U\right>- \left<
B|T_{a}^{\ b}|B\right>)= \lim_{r\to\infty}( \left< U|T_{a}^{\ b}|U\right>
- \left< B|T_{a}^{\ b}|B\right>)_{unren}\nonumber \\ &=&
\lim_{r\to\infty} 2\int_{0}^{\infty}\frac{dw}{e^{8\pi Mw} -1} 
T_{a}^{\ b}\left[\stackrel{\rightarrow}{u}_w,
{\stackrel{\rightarrow}{u}}_w^{\ast}\right]\>.\label{uvi}
\ea
In deriving the above expressions we used the fact that the 
differences between unrenormalized and renormalized quantities are
the same. This because the divergences, being ultraviolet, are state
independent, hence the counterterms are the same for every state.
One sees that the basic quantity entering all the expressions is
$T_{ab}[u_w,u^{\ast}_w]$ which using the decomposition \eqns 
(\ref{u1}), (\ref{u2}) can be written as
\be
T_a\,^b[u_w, u^{\ast}_w] = E\left( \begin{array}{cc} -1 & 0 \\0 & 1
\end{array}\right) +
F\left( \begin{array}{cc}  0 & -1 \\1 & 0 \end{array}\right)\>,
\ee
where
\be
E=\frac{1}{8\pi wf}\left\{ \left[
w^2|R|^2+\frac{dR}{dr^{\ast}}\frac{dR^{\ast}}{dr^{\ast}}\right]
-{f\over r}\left(R\frac{dR^{\ast}}{dr^{\ast}}+R^{\ast}\frac{dR}{dr^{\ast}}\right)
+|R|^2\frac{f^2}{r^2}\right\} \ee and
\be
F=-\frac{i}{8\pi f}\left( R^{\ast}\frac{dR}{dr^{\ast}}-
R\frac{dR^{\ast}}{dr^{\ast}}\right) \ee with $f \equiv (1-2M/r)$.
Using the asymptotic expansions \eqns (\ref{ho}), (\ref{in})  for
the radial function the limiting behaviours of $\left< T_{a
b}\right>$ can be evaluated. %\\ \noindent

Let us start by discussing what is perhaps the 
most interesting quantity, namely the Hawking flux for this theory, 
whose value has been the object of a lively debate. Only for the Unruh
state is there a nonvanishing component of the flux
$T_{r^{\ast}}^{\ t}$. Note also that the Wronskian contained in
$F$ is constant so it can be calculated for all $r$ from the
asymptotic expansion. We find therefore 
\ba \left< U|T^{\
t}_{r^{\ast}}|U\right>&=& \left< U|T^{\ t}_{r^{\ast}}|U\right> -
\left<B|T^{\ t}_{r^{\ast}}|B\right> \nonumber \\ &=&  (\left< U|T^{\
t}_{r^{\ast}}|U\right>- \left<B|T^{\
t}_{r^{\ast}}|B\right>)_{unren}=f^{-1}\, \dot{E}_U\, ,
\ea  
where
\be
\dot{E}_U=
\frac{1}{2\pi}\int_{0}^{\infty}\frac{wdw}{e^{8\pi Mw}-1}|B(w)|^2\>
\ee
is the energy flux at infinity.  Not surprisingly, this flux is positive;  
i.e., there is no antievaporation of the black hole in this theory.
%So, we need to know the greybody factor $|B(w)|^2$ to calculate the
We can calculate the total flux using  
Page's result \cite{Pa} for the $w\to 0$ 
asymptotics of the greybody factor $|B(w)|^2$ for $l=0$ mode, 
\be
|B(w)|^2=16M^2w^2 \>.
\ee
Integration over the frequencies leads to the approximate Hawking flux 
in this 2D theory: 
\be
\dot{E}_U^{\scriptsize\mbox{Page}}
=\frac{1}{7680\pi M^2} \>.
\ee
This low-frequency approximation for the transmission amplitude 
should work quite well since  high frequencies will not contribute to the
flux  because of the Planckian exponent.
Note that the value of the Hawking flux 
$\dot{E}_U^{\scriptsize\mbox{Page}}$
is exactly $1/10$ of the
corresponding value coming from the Polyakov theory (massless minimally
coupled 2D scalar field). 
This damping is due to the potential barrier present in the radial
equation (\ref{barr}) which reflects the coupling of the scalar field
with the dilaton.
In the Polyakov theory there is no potential barrier, hence  
$|B(w)|^2 \equiv 1$ and 
$\dot{E}_U^{\scriptsize\mbox{Polyakov}}
=10 \dot{E}_U^{\scriptsize\mbox{Page}}
$.

Accurate numerical
calculations of the greybody factor for $l=0$ mode and the 
corresponding Hawking flux give   
\be
\dot{E}_U^{\scriptsize\mbox{numerical}}
=C \, \dot{E}_U^{\scriptsize\mbox{Page}} \> ,
\ee
where the coefficient 
\be
C\approx 1.62 \>.
\ee
It is interesting to compare the $2D$ (s-mode) 
Hawking flux with that of the $4D$ black hole. B.S.~DeWitt \cite{De}
provides an approximate formula for the transmission coefficient, 
$|B(w)|^2=27M^2w^2$, which takes into account the
contribution to the $4D$ Hawking flux of all momenta (this gives 
$C=1.69$),
whereas numerical calculations \cite{Elst} of the $4D$ Hawking flux at
infinity
give
$\dot{E}_U^{\scriptsize{4D\mbox{-numerical}}} 
\approx 1.79 \dot{E}_U^{\scriptsize{Page}} $ 

Using the asymptotic expansion we can extract the leading behaviour of
$\left< U| T_{a}^{\ b}|U\right>$ near the horizon and at infinity
(see \eqns (\ref{uv}), (\ref{uvi})): 
\be
\left< U|T^{\,b}_a|U\right>_{r\to 2M}\sim \frac{1}{7680 \pi M^2
}\(
\begin{array}{cc}
1/f & -1 \\ 1/f^2 & -1/f \end{array} \) \> , \label{cd}
\ee 
and
\be
\left<U| T^{\,b}_a|U\right>_{r \to \infty}\sim\frac{1}{7680 \pi
M^2 }\(
\begin{array}{cc}
-1 & -1 \\ 1 & 1 \end{array} \)  \> , 
\ee 
where now $a,b=r,t$. From \eqn  
(\ref{cd}) one sees the negative energy flux entering the black
hole horizon which compensates the Hawking radiation at infinity.
%\\ \noindent

Using similar methods one obtains (see \eqns (\ref{nb}), (\ref{vv})) 
\be
\left< B|T^{\,b}_a|B\right>_{r\to 2M}\sim\frac{1}{384\pi M^2 f} \(
\begin{array}{cc}
1  & 0 \\ 0 & -1  \end{array} \) \label{lbn}\ee and
\be \label{ciao}
\left< H|T^{\,b}_a|H\right>_{r\to\infty}\sim \frac{1}{384\pi M^2 }\(
\begin{array}{cc}
-1  & 0 \\ 0  & 1  \end{array} \)\label{lbl} \>. \ee This last
equation shows clearly that the Hartle-Hawking state
asymptotically describes a thermal bath of 2D radiation at the
Hawking temperature $T_H=(8\pi M)^{-1}$. The prefactor is the
expected $\frac{\pi}{6}T_H^2$. This is indeed the leading
contribution (in a $1/r$ expansion) for the s-mode in flat space
(see Appendix). 

\setcounter{equation}{0}
\section{$\left< T_{a b}\right>$: analytical approximations for the
Boulware
and Hartle-Hawking states}

To obtain an analytical expression for $\left< T_{ab}\right>$
valid for every $r$ ($2M<r<\infty$) we use point-splitting
regularization followed by a WKB approximation for the modes. The
renormalized expression $\left< T_{a b}\right>$ is then obtained
by subtraction of renormalization counterterms 
$\left<T_{ab}\right>_{DS}$ coming from the DeWitt-Schwinger expansion 
of the Feynman Green function and removal of the regulator (point
separation). This method is nicely explained in the seminal work of
Anderson et al. \cite{AnHiSa}
on $\left< T_{\mu\nu}\right>$ in spherically
symmetric static spacetimes, to which we refer the reader for all
details. This section is just an application of their general
method to our (much simpler) s-wave case. Here we just outline the
main points of the derivation. %\\ \noindent

One first analytically continues
the spacetime metric into an Euclidean form by letting $\tau=it$: 
\be
ds^2=fd\tau^2 +f^{-1}dr^2\>. 
\ee 
By the point-splitting method
$\left< T_{a b}\right>_{\scriptsize{\mbox{unren}}}$ is calculated by taking
derivatives of the quantity 
$\left<\varphi(x)\varphi(x^{\prime})\right>$ and then letting
$x^{\prime}\to x$. When the points are separated one can show that
\be
\left< T_{a b}\right>_{\scriptsize{\mbox{unren}}} 
  =  e^{-\left(\phi(x)+\phi(x^{\prime})\right)}
\left[\frac{1}{2}\left( g^{{c}^{\prime}}_{a}G_{E;{c}^{\prime}b}+
g_{b}^{{c}^{\prime}}G_{E;a{c}^{\prime}}\right) 
- \frac{1}{2}g_{a b}g^{c{d}^{\prime}} 
G_{E;c{d}^{\prime}} \right] \label{timunu} \>,
\ee
where $G_E$ is the Euclidean Green function satisfying the
equation
\be
\nabla^a ( e^{-2\phi}\nabla_a G_E \left( x,x^{\prime}\right) )= -
g^{-1/2}(x){\delta}^2 \left( x,x^{\prime}\right) \> , \label{inac}
\ee
and the quantities $g^{{c}^{\prime}}_{a}$ are the bivectors of
parallel transport. The integral representation for
$G_E(x,x^{\prime})$ used by Anderson et al. \cite{AnHiSa}  is the
following: 
\be
G_E \left( x,x^{\prime}\right)= \int d\mu \cos\left[\omega\left(
\tau -{\tau}^{\prime}\right) \right] p_{\omega }\left(
r_{<}\right) q_{\omega }\left( r_{>}\right) \> , \label{gcv}
\ee 
where,
for an arbitrary function $F$,
\[ \int d\mu F\left( \omega \right)\equiv
\frac{1}{4{\pi}}\int_0^{\infty}d\omega\, F\left(\omega \right)
\] if $T=0$ (Boulware state), whereas for $T>0$
\[ \int d\mu F\left( \omega
\right)\equiv 2T\sum_{n=1}^{\infty}F\left(\omega_n\right)+ T F\left(
0\right)\]
and $\omega_n = 2\pi n T$ . %\\ \noindent 

The modes $p_{\o}$ and $q_{\o}$ are analogous to the
radial functions ${\stackrel{\leftarrow}{R}}/r$,
${\stackrel{\rightarrow}{R}}/r$ used in the previous section. They
satisfy the Euclidean version of \eqn (\ref{barr}), which we write
as
\be
f\f{d^2 S}{dr^2}+ \f{2}{r}\left(1 -
\f{M}{r}\right)\f{dS}{dr}-\f{{\o}^2}{f}S=0 \, , \label{moeq}
\ee 
and the Wronskian condition 
\be C_{\o}\left[
p_{\o}\f{dq_{\o}}{dr}-q_{\o}\f{dp_{\o}}{dr}\right]=-\f{1}{fr^2}\>.
\ee 
To express these modes we use the WKB approximation: 
\ba 
 & &  p_{\o}\equiv \f{1}{r\sqrt{2 W(r)}}\exp\ls\int^r\f{W(r)}{f}\, dr\rs
      \>, \nonumber \\ 
 & &  q_{\o}\equiv \f{1}{r\sqrt{2 W(r)}}\exp\ls
      -\int^r\f{W(r)}{f}\, dr\rs \>.\label{mow}
\ea 
By this change of variables one sees that the Wronskian 
condition is satisfied by $C_{\o}=1$. 
Substituting \eqns (\ref{mow}) into the mode
equation \eqn (\ref{moeq}) one finds that the function $W(r)$ has
to satisfy
\be
W^2={\o}^2+V+\f{f}{2W}\ls
f\f{d^2W}{dr^2}+\f{df}{dr}\f{dW}{dr}-\f{3f}{2W}\(\f{dW}{dr}\)^2\rs
\label{W2}
\ee 
where $V=\f{f}{r}\f{df}{dr}$. This is solved iteratively
starting from the zeroth-order solution
\be
W = {\o}\>. 
\ee 
By this method one obtains an explicit form for
the modes $p_w, q_w$ to be inserted in the general expression of
$G_E$ (\eqn (\ref{gcv})). Taking derivatives of the latter
quantity as indicated in \eqn (\ref{timunu}) one eventually arrives
at the following expression for $\left< T_a^{\ b}\right>_{unren}$: 
\begin{eqnarray}
\left< T_t\,^t\right>_{unren} &=& -\left< T_r\,^r\right>_{unren} =
 e^{-2\phi}\int d\mu \cos \( \o {\epsilon}_{\tau} \) \ls
-\f{1}{2} g^{tt^{\prime}} {\o}^{2} {A}_{1} -
\f{1}{2}g^{rr^{\prime}} A_2 \rs \nonumber \\ & + &  e^{-2\phi} i
\int d\mu \o \sin \( \o\et \) \ls
-\f{1}{2}g^{rt^{\prime}}A_3-\f{1}{2}g^{tr^{\prime}}A_4\rs  \>,
\label{unren}
\end{eqnarray}
where
\begin{eqnarray*} A_1=p_{\o}q_{\o}
~,\hskip 1cm A_2=\f{dp_{\o}}{dr}\f{dq_{\o}}{dr} 
~,\hskip 1cm A_3=q_{\o}\f{dp_{\o}}{dr}
~,\hskip 1cm A_4=p_{\o}\f{dq_{\o}}{dr}~,
\end{eqnarray*}
and ${\epsilon}_{\tau} \equiv \tau - \tau^{\prime}$. For the sake of
convenience the points are split in time only so that $r^{\prime}=r$. 
%\\ \noindent

The expansion for the bivectors is 
\be
g^{tt^{\prime}}=-\frac{1}{f}-\frac{f'^2}{8f}\epsilon^2
+O(\epsilon^4)\>, 
\ee 
\be
g^{tr^{\prime}}=-g^{r^{\prime}t}=-\frac{f'}{2}\epsilon+O(\epsilon^3)\>,
\ee
\be g^{rr^{\prime}}= f+\frac{f'^2 f}{8}\epsilon^2
+O(\epsilon^4)\>,
\ee 
where $f'\equiv df/dr$. %\\ \noindent

Eventually one arrives at the following expression for 
$\left<T_t^t\right>_{unren}$ in the zero temperature case: 
\be
\left< B|T_t^{\ t}|B\right>=-\left< B|T_r^{\ r}|B\right>=
\frac{1}{2\pi f}\left[ \frac{1}{\epsilon ^2}+\frac{M^2}{2r^4} +
\frac{f^2}{4r^2}(2\gamma +\ln(4\lambda^2\epsilon^2) ) \right] \>, 
\ee 
which shows $1/\epsilon^2$ and $\ln\epsilon$ divergences as $\epsilon\to 0$
($\lambda$ is a lower limit cutoff in the integral over $\omega$ and
$\gamma$ is the Euler constant).
To obtain the renormalized expressions one needs to subtract from
the above expressions the renormalization counterterm 
$\left<T_a^{\ b}\right>_{DS}$ obtained using the following Green 
function (see \cite{BuChFu} for the details): 
\be
G^{(1)}(x,x')=\frac{e^{\phi(x) + \phi(x')}}{2\pi}[-(\gamma
+\frac{1}{2}\ln(\frac{m^2\sigma}{2})
)(1+(\frac{R}{12}-\frac{a_1}{2})\sigma) +\frac{a_1}{2m^2} +
...]\> , \label{reg} 
\ee 
where $m^2$ is an infrared cutoff and $a_1$ is the DeWitt-Schwinger
coefficient for the action (\ref{act}), 
\be{a_1}=\frac{1}{6}(R - 6\( \nabla\phi\)^2 + 6\Box\phi)\>.
\ee 
Here $R$ is the Ricci scalar for the 2D metric 
and $\sigma$ is one-half of the square of the distance between the
points $x$ and $x^{\prime}$ along the shortest geodesic connecting
them. For our splitting 
\ba 
 & & \sigma^{t}=\sigma^{; t}=\epsilon
     +\frac{f'^2}{24}\epsilon^3 +O(\epsilon^5)\>,\nonumber \\ 
 & & \sigma^{r}=\sigma^{; r}=-\frac{f'f}{4}\epsilon^2
     +O(\epsilon^4)\>,
\ea 
and $\sigma=\sigma^{a}\sigma_{a}/2$. This
allows the counterterm to be evaluated in an $\epsilon$ expansion: 
\ba \left< T_t^{\ t}\right> _{DS}&=&\frac{1}{2\pi
f}\left[\frac{1}{\epsilon^2}
+\frac{5}{12}\frac{M^2}{r^4}+\frac{1}{6}\frac{fM}{r^3}
+\frac{f^2}{4r^2}(2\gamma +\ln(m^2\epsilon^2f) )\right]\>, \nonumber \\
\left<
T_r^{\ r}\right>_{DS}&=&\frac{1}{2\pi
f}\left[-\frac{1}{\epsilon^2}-\frac{5}{12}\frac{M^2}{r^4}
+\frac{fM}{2r^3}-\frac{f^2}{4r^2}(2\gamma 
+\ln(m^2\epsilon^2f) )\right]\>.\label{cicco}
\ea The renormalized expectation
value is then defined as
\be
\left< T_{a b}\right>=Re\left[ \lim_{\epsilon\to 0} (\left< T_{a
b}\right>_{unren} - \left< T_{a b}\right>_{DS})\right]\>.\ee In
the Boulware state this yields
\be
\left<B| T_t\,^t|B\right>_{WKB}=\f{1}{2\pi f}\(
\f{1}{12}\f{M^2}{r^4}-\f{1}{6}\f{fM}{r^3}
-\frac{f^2}{4r^2}\ln(\frac{m^2 f}{4\lambda^2}) \)\>,\label{vel}
\ee
\be
\left<B| T_r\,^r|B\right>_{WKB}=\f{1}{2\pi f}\( -
\f{1}{12}\f{M^2}{r^4}-\f{1}{2}\f{fM}{r^3}
+\frac{f^2}{4r^2}\ln(\frac{m^2f}{4\lambda^2}) \)\>.\label{vem} \ee
Note that $\left< B|T_{a b}|B\right>$ has the correct trace
anomaly:  
\be
\left< B|T^a_a|B\right>_{WKB} =\frac{a_1}{4\pi}=
\frac{1}{24\pi}(R-6(\nabla\phi)^2+6\Box\phi)
=-{1\over 24\pi}\left({d^2 f\over dr^2}+{6\over r}{df\over dr}\right)
=-{M\over 3\pi r^3}\>.\ee
 \par \noindent
It is easy to show
that $\left< B|T_{a b}|B\right>$ is not conserved.
Reparametrization invariance of the action (\ref{act}) gives the
following nonconservation equation ($\!$\cite{BaFa}, \cite{KuVa})
\be
\nabla_a \left< T^a_b \right>=-\frac{1}{\sqrt{-g}}\left< \frac{\delta
S}{\delta\phi}\nabla_b\phi\> \right>.\label{coe}
\ee 
A ``source term'' is present because of the coupling with the dilaton. 
\Eqns (\ref{coe}) are nothing but the 4D conservation equations 
$\nabla_{\mu}\left<T^{(4)\mu}_{\nu}\right>=0$ for the minimally 
coupled massless scalar field of the action (\ref{foact}).  This 
allows us to define a ``pressure'' for our 2D model by rewriting 
\eqns (\ref{coe}) as  
\ba 
 & & 8\pi r T^{\theta}_{\theta}= {\partial}_r
     T^r\,_r +\f{M}{r^2 f}\(T^r\,_r - T^t\,_t\) \>, \nonumber \\ 
 & & {\partial}_{r}T^{r}_{t}=0 \>. \label{cep} 
\ea 
Then from \eqns (\ref{vel}), (\ref{vem}) and (\ref{cep}) one has
\be
\left< B|T^{\theta}_{\theta}|B\right>=\frac{1}{64\pi^2}\left[
\frac{8M}{r^5}
-\frac{2}{r^4}(1-\frac{4M}{r})\ln(\frac{m^2f}{4\lambda^2})\right]\>.
\ee 
It is rather interesting to note that provided we set $m=2\lambda$ the 
above expressions for $\left<B| T_{a}^{\ b}|B\right>$ and the pressure
coincide exactly with the ones derived from the ``anomaly
induced'' effective action for the theory (\ref{act})
\cite{BaFa}. 

The thermal case is treated similarly. Evaluating the
sum over $n$ using the Plana sum formula, one finds that the stress
tensor at finite temperature is obtained from the zero-temperature
one by making the substitution
\be
\ln(\frac{m^2f}{4\lambda^2}) \to \left\{ 2\gamma +
\ln(\frac{m^2\beta^2f}{16\pi^2})\right\} \ee 
%($\gamma$ is Euler constant) 
and adding the traceless pure radiation term
\be
(T_t^{\ t})_{rad}=-(T_r^{\ r})_{rad}=-\frac{\pi}{6\beta^2 f} \> ,  
\ee
where $\beta=T^{-1}$ . %\\ \noindent

Summarizing, we find that in the WKB 
approximation for the Hartle-Hawking state
\be \label{nono}
\left<H|
T_t\,^t|H\right>_{WKB}=-\frac{\pi}{6\beta^2 f}+ \f{1}{2\pi f}\left[
\f{1}{12}\f{M^2}{r^4}-\f{1}{6}\f{fM}{r^3}
-\frac{f^2}{4r^2}\(2\gamma + \ln(\frac{m^2 \beta^2f}{16\pi^2})\) \right]
\>, \label{vec} 
\ee
\be \label{nonno}
\left<H| T_r\,^r|H\right>_{WKB}=\frac{\pi}{6\beta^2 f}+ \f{1}{2\pi
f}\left[ - \f{1}{12}\f{M^2}{r^4}-\f{1}{2}\f{fM}{r^3}
+\frac{f^2}{4r^2}\(2\gamma +\ln(\frac{m^2\beta^2f}{16\pi^2})\) \right]\>,
\label{ved} 
\ee
\be
\left< H|T^a_{a}|H\right>_{WKB}=\left< B|T^a_a|B\right>_{WKB} =
-{M\over 3\pi r^3}, \label{Taa} 
\ee
\be
\left< H|P|H\right>_{WKB} =\frac{1}{64\pi^2}\left[ \frac{8M}{r^5}
-\frac{2}{r^4}(1-\frac{4M}{r})\(2\gamma 
+ \ln(\frac{m^2\beta^2 f}{16\pi^2})\)\right]\>,
\ee 
where in this case $\beta=T_H^{-1}$. %\\ \noindent

The analytic
expressions we have obtained for $\left< B|T_{a b}|B\right>_{WKB}$
and $\left< H|T_{a b}|H\right>_{WKB}$ have the correct asymptotic
behaviours at $r\to\infty$ as inferred in the previous section. 
$\left< B|T_{a}^{\ b}|B\right>_{WKB}$ does indeed have the limiting 
form \eqn (\ref{lbn}) as the horizon is approached, whereas 
$\left< H|T_{a}^{\ b}|H\right>_{WKB}$ for large $r$ describes 
thermal radiation at the Hawking temperature in agreement with 
\eqn (\ref{lbl}).

In the Hartle-Hawking state the stress tensor 
should be regular on the horizon.
This means that on the horizon the leading term of 
$\left< H|T_{a}^{\ b}|H\right>$ should be proportional to the $2D$ metric,
since the manifold of the Euclidean instanton is regular 
and the Hartle-Hawking state respects all its symmetries.
But the trace of the stress tensor is known exactly because we
know the conformal anomaly (\ref{Taa}) in $2D$. 
So, on the horizon we should obtain
\be
\left< H|T_{a}^{\ b}|H\right>\Big|_{r=2M}={1\over 2}\delta_a^b~\left<
H|T^c_c|H\right>\Big|_{r=2M}=-\frac{1}{48\pi M^2}~\delta_a^b~.
\ee
In the vicinity of the horizon 
this provides only the leading term.
Our results \eqns (\ref{nono}), (\ref{nonno}) fulfill this condition.
However, to ensure 
finiteness of the stress tensor near the horizon in a regular frame 
one should satisfy the stronger condition
\be
{\left< H|T_{t}^{t}|H\right>-\left< H|T_{r}^{r}|H\right>\over
f}={\mathrm {finite}}\>.
\ee 
This leads to serious concerns regarding the expression we
found for the Hartle-Hawking state using the WKB approximation. 
The logarithmic term present in \eqns (\ref{vec}), (\ref{ved})
causes $\left< H|T_a^{\ b}|H\right>_{WKB}$ to be logarithmically 
divergent at the horizon when calculated in a free-falling frame.
This kind of logarithmic divergence is also present in the 4D
calculation of Anderson et al. for non-vacuum spacetimes like
Reissner-Nordstr\"om \cite{AnHiSa}. However, numerical computations
performed by
the same authors give no indication that this divergence actually
exists. Similarly, we suspect that the log term we have in \eqns
(\ref{vec}), (\ref{ved}) is an artifact of the WKB approximation
which, as we shall see in the next section, breaks down near the
horizon.

\setcounter{equation}{0}
\section{$\left< H|T_{a}^{\ b}|H\right>$ near the horizon}

From the discussion of the previous section one can see the
disappointing fact that in the Hartle-Hawking state the energy
density as measured by a free-falling observer in the WKB
approximation diverges
logarithmically as one approaches the horizon $r=2M$. On physical
grounds we do not expect this to happen, since the Hartle-Hawking
state is defined in terms of modes which are regular at the horizon. The
origin of the log term in $\left< H|T_a^{\ b}|H\right>_{WKB}$ is
in the counterterms $\left< T_a^{\ b}\right>_{DS}$ (see \eqn 
(\ref{cicco})). The WKB approximation for the modes produces in
$\left< T_a^{\ b}\right>_{unren}$, besides terms of the form
$\ln\epsilon$ and and $1/\epsilon^2$ which are cancelled by the
counterterms, only a monomial involving $f$ and powers of $r$. The
natural question which arises is whether one can trust the WKB
approximation near the horizon. %\\ \noindent

The Euclidean modes $~Y=\left(~rp_{\omega},~rq_{\omega}\right)~$ 
(see \eqn (\ref{mow}))
satisfy a Schr\"odinger-like equation 
\be
{d^2 Y\over {dr^*}^2}-U(r^*)Y=0~, \hskip 0.6cm
U(r^*(r))=\omega^2+V~,\hskip 0.6cm 
V={2M\over r^3}f~,\hskip 0.6cm f=\left(1-{2M\over r}\right)~.
\ee 
Solving iteratively the equation for $W^2$ (see  \eqn (\ref{W2})), 
\be
W^2=\omega^2+V+{1\over{4 W^2}}{d^2(W^2)\over {dr^*}^2}-{5\over16
~W^4}\left({d(W^2)\over {dr^*}}\right)^2 \>, 
\ee
we get
\begin{eqnarray}
W^2&=&(W^2)_0+(W^2)_1+(W^2)_2+\dots \>, \\
(W^2)_0&=&\omega^2 \>, \\
(W^2)_1&=&V \>, \\
(W^2)_2&=&{1\over4(\omega^2+V)}{d^2V\over {dr^*}^2}-
{5\over16~(\omega^2+V)^2}\left({dV\over {dr^*}}\right)^2 \>, \\
\end{eqnarray}
Note that $V\sim f$, as do all its derivatives 
$\partial_{r^*}^k V$. For $\omega=0$ the first terms $(W^2)_0$ 
and $(W^2)_1$ vanish at the horizon
while the next ``correction'' $(W^2)_2$ is already finite.
This indicates that the WKB approximation can not work near 
the horizon for the zero-frequency mode. 
For the modes with non-zero $\omega=\omega_n=(4M)^{-1}n$ we have 
\be
W^2={1\over (2 M)^2}\left[{1\over 4}n^2
+f\left(1+{1\over n^2}\right)+O(n^{-4})\right]+ O(f^2) \> .
\ee
One can see that the convergence of the WKB series implies 
that $n$ is at least greater then 1.
Evaluation of the corresponding series 
for $\left<\hat{\varphi}^2\right>$
and the stress tensor $\left< H|T_{a}^{\ b}|H\right>$ near the horizon
leads to exactly the same conclusion: 
\be n\gg 1 \>.\ee

Clearly, the standard WKB approximation can not
be applied for the calculation of the contribution of the 
$n=0$ and $n=1$ modes
to quantum averages near the horizon. 
To obtain a more reliable analytical
expression for $\left< H|T_a^{\ b}|H\right>$ near the horizon we
need a better  
approximation for the Green function
for these modes. %\\ \noindent 

In Ref. \cite{TomimatsuKoyama}  it was demonstrated that 
a more accurate calculation of the contribution of  the $n=0$ mode  
cures the analogous logarithmic divergence in the total
$\left<\hat{\varphi}^2\right>_{WKB}$.  
Here we follow a similar approach to analyze the stress tensor
(see also \cite{RoSt}).
%\\ \noindent

One can decompose the thermal Euclidean Green function 
for the $~Y$ modes as
\be
 G_E(\tau, r; \tau^{\prime},r^{\prime})
  =  \frac{1}{\beta} \sum_{n=-\infty}^{+\infty} 
     \frac{\cos w_n(\tau -\tau^{\prime})}{[f(r) f(r^{\prime})]^{1/4}} 
     \, G_n(r,r^{\prime}) \>,
\ee 
where we write $w_n$ for the
frequency instead of just $w$ as before to make the dependence on
$n$ more clear ($w_n=2\pi n/\beta$). %\\ \noindent

Near the horizon the function
$G_n(r,r^{\prime})$ satisfies the following differential equation
(with $r\neq r^{\prime}$): 
\be
\partial_L^2 G_n -\left( \frac{\alpha^2}{M^2} +\frac{4n^2-1}{4L^2}
+O(f)\right) G_n=0 \label{digi} \>, 
\ee 
where $L$ is defined by
\be
dL=\frac{dr}{f^{1/2}} 
\ee 
and 
\be
\alpha^2=\frac{1}{6}+\frac{n^2}{12}\>.
\ee 
The differential equation (\ref{digi}) admits solutions in terms of Bessel
functions of imaginary argument, 
\be
G_n(r,r^{\prime})=(LL^{\prime})^{1/2}I_n(\frac{\alpha L_{<}}{M})
K_n(\frac{\alpha L_{>}}{M})\>. \ee 
One can show that this solution obeys the
derivative condition resulting from integrating the differential
equation (\ref{inac}) for $G_E$ across the delta function
singularity at $\tau=\tau^{\prime}, r=r^{\prime}$. Using the above
Green function one can calculate the corresponding contribution to
the stress tensor for each $n$ near the horizon. %\\ \noindent

For a contribution to the Green function of the form
\be
e^{-iw_n(t-t^{\prime})}F_n(r,r^{\prime})\ee the corresponding
contribution to the unrenormalized stress tensor in the
Hartle-Hawking state is
\be
\left< T_a ^{\ b}\right>_n= \lim_{r\to r^{\prime}} \left\{
-\frac{f}{2r^2}\left[ 1-r(\partial_r +\partial_{r^{\prime}})
+r^2\partial_r\partial_{r^{\prime}} \right]
+\frac{w_n^2}{2f}\right\}F_n(r,r^{\prime})\left( \begin{array}{cc}
1 & 0 \\0 & -1 \end{array}\right)\>.\ee For the $n=0,1,2$ modes
one obtains
\be
\left< T_a ^{\ b}\right>_0=\left[ \frac{7f}{240\pi M^2}
+O(f^2)\right]\left( \begin{array}{cc} 1 & 0 \\0 & -1
\end{array}\right)\>,\ee
\be
\left< T_a ^{\ b}\right>_1= \frac{1}{64\pi M^2}\left[ \frac{1}{f}
+f(2\gamma +\ln f) -\frac{f}{3} +O(f^2) \right] \left(
\begin{array}{cc} 1 & 0 \\0 & -1
\end{array}\right)\>,\ee
\be
\left< T_a ^{\ b}\right>_2=\left[ \frac{1}{32\pi M^2
f}-\frac{f}{48\pi M^2} +O(f^2)\right] \left(
\begin{array}{cc} 1 & 0 \\0 & -1
\end{array}\right)\>.\ee
Note that each $n>0$ contribution should be double-counted 
to account for the $n<0$ modes as well. 

These results should be compared to those
coming from the WKB approximation. The $n=0$ mode does not make any
contribution to $\left< T_{a}^{\ b}\right>_{WKB}$ whereas the
contribution of an individual mode with $n\neq 0$ is 
\be
\left< T_{a}^{\ b}\right>_{WKB~n}= \left[\frac{|n|}{64\pi
M^2f}-\frac{f}{32\pi
|n| M^2}
+O(f^2)\right]\left( 
\begin{array}{cc} 1 & 0 \\0 & -1
\end{array}\right)\>.
\ee
Taking the difference we find the correction
to $\left< H|T_{a b}|H\right>_{WKB}$ due to the first three 
modes to be 
\be
\delta \left< T_a^{\ b}\right> _{n=0,\pm 1,\pm 2} = \left[
\frac{f}{32\pi M^2}(2\gamma + \ln f) +\frac{17 f}{240\pi
M^2}\right]\left(
\begin{array}{cc} 1 & 0 \\0 & -1
\end{array}\right) +O(f^2)\>.\label{cicp}\ee
Comparing this with \eqns (\ref{vec}), (\ref{ved}) we find that the
corrections above exactly cancel the logarithmic term at the event
horizon to order $f\ln f$. Only the $n=\pm 1$ modes contribute
such terms. For $|n|>1$ only higher-order log terms (i.e. $f^2\ln
f$ etc.) are produced which will cause no divergence. Proceeding
in a similar way we find the correction to the pressure, 
\be 
\delta
P_{n=0,\pm 1,\pm 2}=\frac{1}{16\pi M^2}\left[ -\frac{83}{960\pi
M^2} -\frac{1}{32\pi M^2}(2\gamma +\ln f) +O(f)\right]
\>.\label{blob}
\ee 
Again this cancels exactly the log term in
$\left< H|P|H\right>_{WKB} $. We can therefore conclude that for
our 2D theory \eqn (\ref{act}) the $\left< H|T_a^{\ b}|H\right>$
and $\left< H|P|H\right>$ are regular (in a free-falling frame) on
the horizon as expected. The logarithmic term appearing in
$\left< H|T_a^{\ b}|H\right>_{WKB}$ 
is an artifact of the WKB approximation which breaks down for the 
low-$n$ modes near the horizon.  
Furthermore,  the nonlogarithmic terms in \eqn (\ref{cicp}) are of 
order $f$ so we can obtain from \eqns (\ref{vec}), (\ref{ved}) 
the following limiting values for $\left< H|T_a^{\ b}|H\right>$ 
on the horizon:  
 \be
\left< H|T_t^{\ t}|H\right>_{r=2M}= 
\left< H|T_r^{\ r}|H\right>_{r=2M}=-\frac{1}{48\pi M^2}\>.
\ee 
On
the other hand the value of the pressure changes because of the
first term in \eqn (\ref{blob}):  
\be 
\left< H|P|H\right>_{r=2M}=
\frac{1}{64\pi^2}\left[- \frac{23}{40M^4}
+\frac{1}{8M^4}\ln\frac{m^2\beta^2}{16\pi^2}\right] \>.
\ee

\setcounter{equation}{0}
\section{Conclusions}

The main purpose of this paper was to shed some light on the rather
controversial literature existing on the Hawking effect for the
dilaton gravity theory described by the action (\ref{act}). We found
that the Hawking flux is manifestly positive, reduced by a greybody
factor with respect to the corresponding value one gets from the
Polyakov theory (no dilaton coupling). We also showed that the
Hartle-Hawking state corresponds to thermal equilibrium at the
Hawking temperature and that asymptotically ($r\to\infty$) the stress
tensor describes a gas of 2D photons. The regularity of this stress
tensor on the horizon has been proved by a careful expansion of the
Green function in that region eliminating the unphysical logarithmic
divergence predicted by the WKB approximation . One can hope that the
analogous logarithmic WKB divergence appearing in nonvacuum 4D
spacetime can be handled in a similar way. %\\ \noindent 

The analytic
expression for $\left< T_{a}^{\ b}\right>$ we found in section 3 can
be exactly reproduced by the high-frequency approximation for the
effective action in static spacetimes developed by Frolov et al.
\cite{FrSuZe}. This point and the generalization of our
work to arbitrary curvature coupling and mass for the scalar field
will be discussed elsewhere.  %\\ \noindent 

The feature which makes
the theory  (\ref{act}) so attractive is its connection with the 4D
action (\ref{foact}). What can be inferred of the physical 4D theory
from the quantization of the dimensionally reduced theory we have
performed? It is often said that the spherically symmetric reduced
theory should describe the s-wave sector of the higher-dimensional
one. Unfortunately in quantum field theory things are not so easy.
Let us compare the value we found for the energy density in the
Hartle-Hawking state on the horizon with the corresponding value
coming from the quantization of the 4D theory of \eqn (\ref{foact}).
Our result (which should be divided by $4\pi r^2$ to restore 
four-dimensionality) yields the following prediction for the s-wave
contribution to the 4D theory:  
\be 
\left< H|T_t^{(s)\
t}|H\right>_{r=2M}=-\frac{1}{768\pi^2 M^4}\>.\label{alamb}
\ee 
The
value found by Anderson et al. \cite{AnHiSa} quantizing the 4D theory
is 
\be \left< H|T_t^{\ t}|H\right> _{r=2M}=\frac{1}{3840\pi^2
M^4}\>.\label{clbn}
\ee 
The discrepancy is striking. Our 2D derived
result is significantly larger than and opposite in sign to the
expected 4D value. One can argue that the value of \eqn (\ref{clbn})
includes the contribution of all $l$ modes and not just the $s$ one.
This might be  true. However it seems unlikely that the
$l>0$ modes should cancel this $l=0$ result \eqn (\ref{alamb}) to a
sufficiently high degree to restore agreement with the 4D stress
tensor. This difference indicates a dismal failure of the dimensional
reduction. But this is not all of the story.  As was shown in
\cite{FrSuZe,Sut}, the $s$-mode contribution to the renormalized
stress-energy tensor of the 4D theory does not coincide with the 
renormalized stress-energy tensor of the 2D reduced theory. The
difference is called the dimensional-reduction anomaly. There is a
suspicion that the actual mismatch between the 2D derived value \eqn
(\ref{alamb}) and the 4D value \eqn (\ref{clbn}) is caused essentially
by this anomaly. A preliminary analysis \cite{Su} seems to confirm
this idea.

\bigskip

{\bf Acknowledgments:~} 
This work was  partly supported  by  the
Natural Sciences and Engineering Research Council of Canada. VF and AZ
are grateful to the Killam Trust for its financial support.

\appendix
\renewcommand{\theequation}{\thesection.\arabic{equation}}

\section{s-mode contribution to the 4D stress tensor in flat space
at finite temperature }\setcounter{equation}{0}

In this appendix we determine  the $l=0$ mode contribution 
to $\left< T_{a}^{\ b}\right>_{\beta}$ in flat space for a minimally
coupled and massless 4D scalar field  in a thermal state at the 
temperature $T=\beta^{-1}$.  For this case we know exactly the 
mode-function solutions $\varphi_w$ of the Klein-Gordon equation 
\be \label{klg} \Box\varphi=0 \>.\ee
Insertion of the spherical decomposition
\be
\varphi=\sum_{w,l,m} \varphi_w (t,r) Y_{lm}(\theta,\phi) \ee
reduces \eqn (\ref{klg}) to 
\be
(-\partial_t^2 +\frac{2}{r}\partial_r+ \partial_r^2
-\frac{l(l+1)}{r^2})\varphi_w=0\>.\ee
For the case of interest ($l=0$) the solutions for $\varphi_w$ are just
the ordinary Fourier modes. Taking into account that $0\le r<\infty$ we
must impose Dirichlet boundary conditions at $r=0$. The correctly
normalized s-modes are then
\be \label{vg}
\varphi_w= \frac{-i}{2\pi r\sqrt{w}}e^{-iwt}\sin(wr)\>,\ee
where $w>0$. 
Decomposition of the field operator $\hat \varphi$ in terms of the modes
$\varphi_w$,  
\be \hat \phi(t,r)=\int_{0}^{\infty} dw [\hat a_w \varphi_w(t,r) +\hat
a_w^{\dagger} \varphi^*_w (t,r) ]\>,\ee
gives the stress tensor expectation values 
\be \label{nbm}
\left< T_{\mu}^{\ \nu}\right>_{\beta} = \int_{0}^{\infty}dw 
\frac{2}{e^{\beta w}-1} T_{\mu}^{\ \nu}[\varphi_w, \varphi_w^*]\>,\ee
where 
\be \label{mcm}
T_{\mu \nu}[\varphi_w,\varphi_w^*]=\frac{1}{2}(\partial_{\mu}\varphi_w
\partial_{\nu}\varphi_w^* +
\partial_{\nu}\varphi_w\partial_{\mu}\varphi_w^*) -\frac{1}{2}g_{\mu
\nu}(g^{\rho\sigma}\partial_{\rho}
\varphi_w\partial_{\sigma}\varphi_w^*)\>.\ee
Inserting (\ref{vg}) into (\ref{mcm}) and performing the integral 
in \eqn (\ref{nbm}) we get
\ba
 \left< T_{\mu}^{\ \nu}\right>_{\beta} &=&  \frac{1}{4\pi r^2}\frac{\pi
T^2}{6} 
\( \begin{array}{cccc} 
 -1 & 0 & 0 & 0 \\ 
  0 & 1 & 0 & 0 \\ 
  0 & 0 & 0 & 0 \\
  0 & 0 & 0 & 0 \end{array} \) + 
 \left(\frac{1}{32\pi^2 r^4} -\frac{T^2}{8r^2
\sinh^2(2\pi Tr)}\right)  \left( \begin{array}{cccc} 
  0 & 0 & 0 & 0 \\
  0 & 0 & 0 & 0 \\
  0 & 0 & -1 & 0 \\
  0 & 0 & 0 & -1 
\end{array}\right)  \nonumber \\ 
&+& \left( \frac{T}{8\pi r^3} \coth(2\pi T r) -\frac{1}{16\pi^2
r^4}\right)
\left( \begin{array}{cccc} 
1 & 0 & 0 & 0 \\ 
0 & -1 & 0 & 0 \\ 
0 & 0 & 1 & 0 \\ 
0 & 0 & 0 & 1 \\
\end{array}\right) \nonumber \\
&-& \frac{1}{16\pi^2 r^4} \ln\{ \frac{\sinh(2\pi Tr)}{2\pi Tr}\} 
\left( \begin{array}{cccc} 
1 & 0 & 0 & 0 \\ 
0 & -1 & 0 & 0 \\ 
0 & 0 & 1 & 0 \\ 
0 & 0 & 0 & 1 \\
\end{array}\right) \>.  
\ea
Multiplying by $4\pi r^2$ and taking the limit $r\to \infty$ we obtain
the result (\ref{ciao}), which describes 2D
thermal radiation at the equilibrium temperature $T=T_H=(8\pi M)^{-1}$.

\end{document}